\documentclass[sigconf,nonacm]{acmart}

\renewcommand\footnotetextcopyrightpermission[1]{}
\usepackage{booktabs}
\usepackage{graphicx}
\usepackage{tikz}
\usetikzlibrary{arrows.meta,positioning}
\usepackage{enumitem}

\title{Pixels for Programs? A Cross-Provider Case Study of Input-Token Accounting for Source Code as Text and Images}

\author{Ronak Bhalgami}
\email{ronak42.b@gmail.com}

\begin{abstract}
Long source-code contexts consume many text tokens, motivating the proposal to
render code as images for vision-language models. Recent work asks whether
models can still solve code tasks after this transformation. We examine a
different systems question: how commercial APIs count the resulting requests. We present a
reproducible measurement case study of provider-reported input tokens for raw source
text and a compact rendered-image representation. The benchmark pairs requests
across five programming languages, nine source lengths from 20 to 2,000 lines,
and 15 available model aliases exposed by Anthropic, OpenAI, and Google Vertex
AI. These aliases collapse to approximately five distinct accounting signatures
and are not independent model replications.
Across 675 complete text/image pairs, aggregate image-to-text ratios are 0.135,
0.194, and 0.242, corresponding to reported input-token reductions of 86.5\%,
80.6\%, and 75.8\%, respectively. These totals conceal materially different
break-even behavior: Anthropic and OpenAI images receive lower counts at every
tested size, while Gemini images require 6.95 times as many tokens at 20 lines
and cross below text only at 200 lines in the aggregate. A targeted audit also
reproduces non-monotonic Gemini image accounting across a page boundary.
This study measures black-box request accounting for one compact rendering
pipeline. It does not measure semantic fidelity, task accuracy, latency,
monetary cost, or coding-agent efficiency. We release the scripts,
revision-pinned corpus specification, raw usage records, validators, and
deterministic analysis needed to reproduce and extend the study.
\end{abstract}


\begin{document}
\maketitle
\begingroup
\renewcommand{\thefootnote}{}
\footnotetext{Artifact repository:
\url{https://github.com/ron-42/code-image-token-accounting}}
\endgroup

\section{Introduction}

Source code is a demanding long-context input. A single repository can contain
far more text than a model can process in one request, and even file-level
workflows repeatedly transmit declarations, implementation details, tests, and
tool output. Code-specialized language models are designed for
long contexts and repository-scale use \cite{roziere_2023_codellama}. In a
commercial API, context is also an accounting concern: reported
input tokens commonly determine quotas, request limits, and part of the bill.

Vision-language APIs create an unusual alternative. Instead of serializing
source as text tokens, a client can render the same source onto one or more
images. A two-dimensional page may carry many characters while the API exposes
a much smaller number of visual input tokens. Pixel language models already
show that rendered text can support linguistic representation
\cite{rust_2023_pixel}, while screenshot and document models recover language
and structure directly from pixels \cite{lee_2023_pix2struct,kim_2022_donut}.
More directly, Glyph renders long text to expand effective context
\cite{cheng_2026_glyph}, and recent code studies evaluate visual compression on
summarization, completion, question answering, and clone detection
\cite{zhang_2026_codeocr,longcodeocr_2026}.

Those results make the representation plausible, but they do not make token
accounting obvious. Providers expose different multimodal request formats,
model families, image-processing policies, and usage fields. A fixed image may
incur a minimum charge, scale with resolution, or be partitioned internally.
Multiple pages may not behave like one larger page. Moreover, an attractive
overall ratio can hide a regime in which images cost more than text. This leads
to a practical measurement question:

\begin{quote}
\emph{For a fixed source corpus and rendering pipeline, how do
provider-reported input-token counts change when code is sent as raw text versus
compact images, and how does that relationship vary with source length?}
\end{quote}

We answer this question with paired black-box measurements. We use five
revision-pinned source files, one each for Python, JavaScript, Rust, Go, and
Java, and slice them into nested prefixes at nine sizes. The image arm first replaces leading
spaces with compact indentation markers and then renders the result as PNG
pages. Both arms append the same one-sentence summarization instruction. We
collect the provider's own input-usage field, preserve raw responses
incrementally, pair successful calls by provider, model, language, and size,
and compute weighted ratios from token totals.

The study makes four contributions:

\begin{enumerate}[leftmargin=*,nosep]
  \item a reproducible protocol and artifact containing 1,350 successful API
  calls and 675 complete text/image observations across three providers, while
  preserving the five-file sampling unit explicitly;
  \item a size-stratified measurement showing immediate reductions for
  Anthropic and OpenAI but a 200-line aggregate crossover for Gemini;
  \item a targeted modality audit demonstrating that reported image usage can
  change non-monotonically at a page boundary; and
  \item an explicit validity boundary separating token accounting from
  information preservation and coding-agent utility.
\end{enumerate}

Here, ``reduction'' means only that the API returned a smaller input-token
count. It does not establish that the
image contains equivalent usable information, that inference consumed less
compute, that the request cost less money, or that a coding agent would perform
better. The contribution is a measurement surface and a set of break-even
observations on which a task-quality study can build.

\section{Background and Related Work}

\subsection{Text represented through pixels}

PIXEL replaces a finite subword vocabulary with visual patches of rendered
text, demonstrating cross-script transfer and robustness while also exposing
quality differences from token-based models \cite{rust_2023_pixel}. Rendering
is not a neutral preprocessing choice: character grouping and spatial
redundancy change both learned representations and the model size needed for
downstream performance \cite{lotz_2023_rendering}. Document-oriented systems
provide complementary evidence. Donut maps document images directly to
structured outputs without a separate OCR pipeline \cite{kim_2022_donut}, and
Pix2Struct learns visual language understanding by parsing web-page screenshots
\cite{lee_2023_pix2struct}. This literature establishes machine readability,
not equivalence between source text and aggressively packed code images.

\subsection{Visual text and code compression}

Glyph treats rendered pages as a context-scaling mechanism and jointly studies
compression and downstream accuracy \cite{cheng_2026_glyph}. VTCBench makes the
quality question explicit by testing retrieval, reasoning, and memory under
vision and text compression \cite{vtcbench_2025}. The closest software-engineering
work is CodeOCR, which evaluates visual code across multiple understanding
tasks and compression levels \cite{zhang_2026_codeocr}. LongCodeOCR extends the
setting to long-context summarization, question answering, and completion
\cite{longcodeocr_2026}. These studies are stronger evidence for usefulness
than token counts alone.

DeepSeek-OCR provides direct reconstruction evidence in a specialized setting.
On English document pages containing 600 to 1,300 text tokens, it reports
about 97\% OCR decoding precision below a 10$\times$ text-to-vision-token
compression ratio. Precision falls to about 60\% near 20$\times$ compression
\cite{wei_2025_deepseek_ocr}. This result shows that high-fidelity optical
compression is possible for a model trained specifically for OCR. It does not
establish the same fidelity for source code or for the commercial model aliases
measured here.

We ask a narrower question. Rather than introducing a new compression method or
comparing downstream accuracy, we treat commercial multimodal APIs as black-box
accounting systems and compare their usage reports across providers and source
sizes. This exposes minimum image costs, crossover points, and page-sensitive
discontinuities that may disappear in a task-level aggregate.

\subsection{Alternative context-efficiency mechanisms}

Textual prompt compression retains selected lexical information rather than
changing modality. LLMLingua performs coarse-to-fine token compression
\cite{jiang_2023_llmlingua}; LongLLMLingua adds question-aware compression,
reordering, and recovery for long contexts \cite{jiang_2024_longllmlingua}.
Prompt caching attacks repeated work instead: attention states for shared
prompt modules can be reused across requests \cite{gim_2023_promptcache}.
Retrieval, caching, textual compression, and visual representation are not
mutually exclusive. Unlike visual conversion, however, the first three can
preserve text-native model interfaces. A realistic coding harness should
compare them on task quality, latency, and total cost rather than selecting a
method from input-token ratios alone.

\section{Study Design}

\subsection{Research questions}

We organize the measurement around four descriptive research questions:

\begin{description}[leftmargin=0pt,style=nextline,nosep]
  \item[RQ1: Aggregate accounting.] What weighted input-token ratio does the
  compact-image pipeline receive from each provider?
  \item[RQ2: Scaling and break-even.] How does the ratio change from 20 to
  2,000 lines, and at what tested size does image input first fall below text?
  \item[RQ3: Observed heterogeneity.] How much does the descriptive result vary
  across model aliases and language/file workloads?
  \item[RQ4: Accounting anomalies.] Do image counts evolve monotonically with
  source length and page count?
\end{description}

These questions concern observed API metadata. They neither posit a provider's
internal tokenizer nor interpret reported tokens as equivalent units of
computation across providers.

\subsection{Corpus and sampling}

The corpus contains one source file per language: CPython's
\url{asyncio/base_events.py}, Lodash's monolithic JavaScript source, the Rust
compiler expression parser, Go's \url{net/http/server.go}, and Guava's
\texttt{LocalCache.java}. Every URL is pinned to a Git commit, and the artifact
records each upstream license. The fetcher saves the first 20, 50, 100, 200,
400, 800, 1,200, 1,600, and 2,000 lines as separate snippets.

The nested prefixes hold earlier content fixed as length grows, which makes
transitions easy to inspect. They are not a random sample. Prefixes from the same file overlap and
must not be treated as independent observations. Language effects are likewise
confounded with file, project, formatting style, and code region. For these
reasons, we report pooled accounting totals without confidence intervals or
significance tests.

\subsection{Compact-image treatment}

The treatment is a two-stage transform. First, each four-space indentation level
is replaced by a leading \texttt{>} marker; an irregular run of $N$ spaces is
replaced by \texttt{\^{}N }. A one-line legend explaining this notation is
prepended. Second, \texttt{pxpipe-proxy} renders the transformed source to one
or more PNG pages. The renderer rejects dropped characters and removes stale
pages before writing a new result.

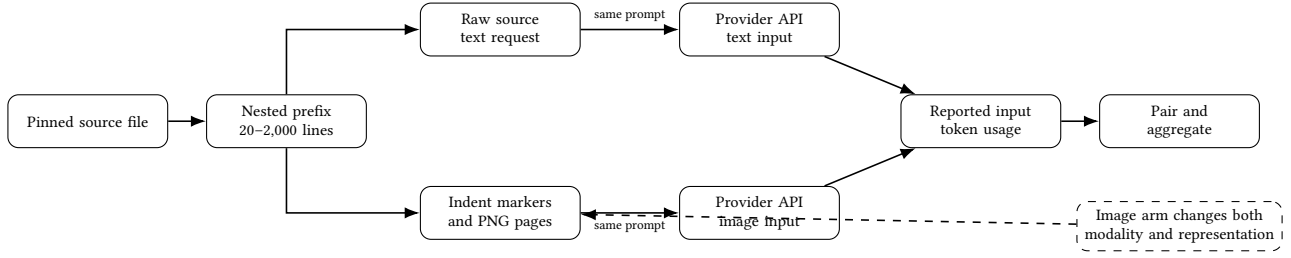
\begin{figure*}[t]
  \centering
  \begin{tikzpicture}[
  node distance=4mm and 5mm,
  box/.style={draw, rounded corners, align=center, minimum height=7mm,
              text width=19mm, font=\scriptsize},
  note/.style={draw, dashed, rounded corners, align=center, text width=25mm,
               font=\scriptsize},
  arrow/.style={-Latex, semithick}
]
\node[box] (source) {Pinned source file};
\node[box, right=of source] (prefix) {Nested prefix\\20--2,000 lines};
\node[box, above right=5mm and 7mm of prefix] (text) {Raw source\\text request};
\node[box, below right=5mm and 7mm of prefix] (compact) {Indent markers\\and PNG pages};
\node[box, right=13mm of text] (textapi) {Provider API\\text input};
\node[box, right=13mm of compact] (imageapi) {Provider API\\image input};
\node[box, below right=5mm and 8mm of textapi] (usage) {Reported input\\token usage};
\node[box, right=of usage] (pair) {Pair and\\aggregate};
\node[note, below=7mm of pair] (confound) {Image arm changes both\\modality and representation};

\draw[arrow] (source) -- (prefix);
\draw[arrow] (prefix) |- (text);
\draw[arrow] (prefix) |- (compact);
\draw[arrow] (text) -- node[above, font=\tiny]{same prompt} (textapi);
\draw[arrow] (compact) -- node[below, font=\tiny]{same prompt} (imageapi);
\draw[arrow] (textapi) -- (usage);
\draw[arrow] (imageapi) -- (usage);
\draw[arrow] (usage) -- (pair);
\draw[arrow, dashed] (confound) -- (compact);
\end{tikzpicture}
  \caption{Paired measurement pipeline. Both arms share the revision-pinned
  source prefix and task instruction, but the compact-image arm additionally
  transforms indentation before rendering. Reported usage measures
  the complete pipeline rather than a modality-only causal effect.}
  \label{fig:pipeline}
\end{figure*}

Figure~\ref{fig:pipeline} shows the paired flow. The raw-text request contains
the untouched snippet and instruction. The image request contains all rendered
pages and the same instruction. This is not a modality-only
experiment: the image arm changes both the transport modality and indentation
representation. We call the treatment \emph{compact image}, not ``the same code
as an image,'' throughout.

\subsection{Providers, requests, and collection}

We queried four Anthropic model aliases, six OpenAI aliases, and five Gemini
aliases during July 2026. Table~\ref{tab:coverage} reports successful calls and
complete pairs for each provider.

\begin{table}[t]
  \centering
  \caption{Measurement coverage. A complete pair contains successful raw-text
  and compact-image requests for the same provider, model, language, and size.}
  \label{tab:coverage}
  \begin{tabular}{lrrr}
\toprule
Provider & Models & Successful calls & Complete pairs \\
\midrule
Anthropic & 4 & 360 & 180 \\
OpenAI & 6 & 540 & 270 \\
Gemini & 5 & 450 & 225 \\
\bottomrule
\end{tabular}

\end{table}

Anthropic requests use the Messages API and one output token. OpenAI requests
use the Responses API, high image detail, and at most 16 output tokens; the
configured reasoning effort is ``none'' except ``low'' for the Codex model.
Gemini requests use Vertex AI \texttt{generate\_content} with at most 16 output
tokens and an eight-second inter-request delay. All variants append exactly:
\emph{``Summarize what this code does in one sentence.''} Output quality is not
scored; generation exists only to produce a valid request and usage response.

Each provider script writes a JSON record after every attempt, supports
resumption without repeating successful calls, and refuses to overwrite an
existing run unless explicitly forced. Anthropic's \texttt{input\_tokens},
OpenAI's input-usage field, and Gemini's prompt-usage field are normalized to
\texttt{input\_tokens}. We do not assume these fields denote the same internal
unit across providers.

\subsection{Metrics and validation}

For each complete pair $i$, the primary ratio is
\[
  r_i = \frac{I_i}{T_i},
\]
where $I_i$ and $T_i$ are provider-reported image and text input tokens.
Values below one favor the compact image under this accounting measure. For an
aggregate group $G$, we report the weighted ratio
\[
  R_G = \frac{\sum_{i \in G} I_i}{\sum_{i \in G} T_i}
\]
and reduction $100(1-R_G)$ percent. This weights each reported token equally
and avoids giving a 20-line request the same influence as a 2,000-line request,
as an arithmetic mean of per-cell ratios would.

The validator rejects duplicate provider/language/model/size/mode keys, unknown
models, nonpositive text counts, and disagreement between raw JSON, paired CSV,
and recomputed ratios. A separate analysis script regenerates every manuscript
table, plot, and numeric-evidence record from the paired CSV files.

\section{Results}

\subsection{RQ1: aggregate accounting}

Table~\ref{tab:overall} gives the weighted result. Across 180 Anthropic pairs,
the compact images receive 0.135 times the raw-text count, a reduction of
86.5\%. Across 270 OpenAI pairs, the ratio is 0.194 and the reduction is 80.6\%.
Across 225 available Gemini pairs, the ratio is 0.242 and the reduction is
75.8\%. These are within-provider descriptive aggregates. They should not be
used to rank providers because model coverage and usage semantics differ.

\begin{table}[t]
  \centering
  \caption{Weighted provider-level accounting. ``Reduction'' is computed from
  summed provider-reported input tokens; it is not a monetary or compute
  saving.}
  \label{tab:overall}
  \begin{tabular}{lrrr}
\toprule
Provider & Complete pairs & Image/text ratio & Input-token reduction \\
\midrule
Anthropic & 180 & 0.135$\times$ & 86.5\% \\
OpenAI & 270 & 0.194$\times$ & 80.6\% \\
Gemini & 225 & 0.242$\times$ & 75.8\% \\
\bottomrule
\end{tabular}

\end{table}

At the scale of the full benchmark, the compact-image arm receives fewer
reported input tokens from all three providers. The differences are large
enough to warrant further evaluation, but these totals do not show whether the
representation is useful or where it becomes advantageous.

\subsection{Aggregation sensitivity}

The weighted statistic answers a volume question: if all measured source
contexts were transmitted once, what fraction of their reported text tokens
would the image requests receive? A second reasonable estimand gives every
provider/model/language/size cell equal weight. Under that unweighted mean,
Anthropic's image/text ratio is 0.186 (81.4\% reduction), OpenAI's is 0.261
(73.9\% reduction), and Gemini's is 1.548 ($-54.8\%$ reduction). Thus Gemini
changes from an apparent 75.8\% reduction under token-volume weighting to an
increase under configuration weighting.

Neither statistic is intrinsically correct for every deployment. The weighted
ratio is appropriate when long contexts contribute most traffic, as they do in
our balanced grid of line counts. The unweighted mean is appropriate when a
random tested configuration is the unit of interest. Their disagreement is
itself a finding: image accounting has strong fixed-cost behavior, so a paper
should publish size-stratified values and state its estimand rather than report
one context-free ``percent saved.'' We use the weighted result as the headline
because this study asks about transmitted context volume, while retaining the
unweighted result as a sensitivity analysis.

\subsection{RQ2: scale and break-even}

\begin{figure*}[t]
  \centering
  \includegraphics[width=0.9\textwidth]{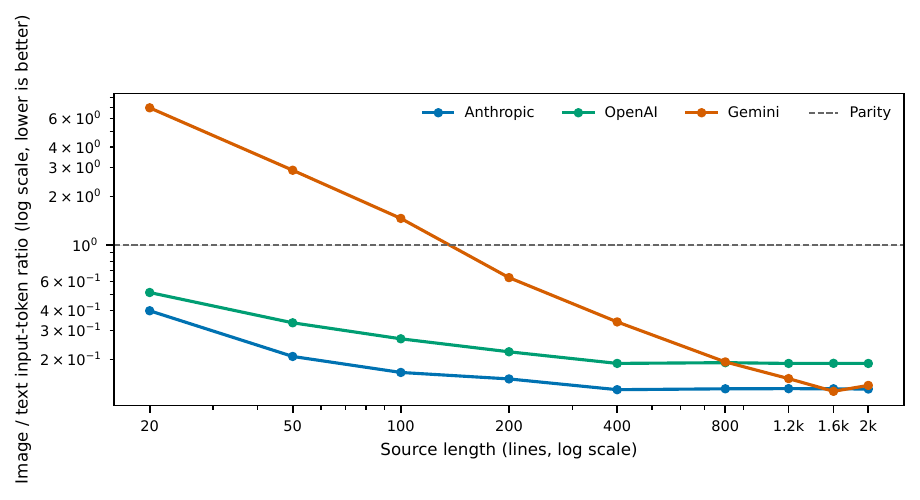}
  \caption{Weighted image/text input-token ratio by source length. Lower is
  better; the dashed line marks parity. Anthropic and OpenAI remain below
  parity, while Gemini's minimum image overhead dominates through 100 lines and
  crosses below text at 200 lines. Points pool available models and five
  language/file prefixes; they are not independent replicates.}
  \label{fig:size}
\end{figure*}

Figure~\ref{fig:size} plots weighted image/text ratios by source length, pooling
models and languages within each provider. The log scale is necessary because
Gemini spans both sides of parity by a wide margin. Anthropic starts at 0.397
for 20 lines and reaches roughly 0.132 at 2,000 lines. OpenAI starts at 0.514
and approaches 0.189. Both are below parity at every tested size.

Gemini behaves differently. Its ratios are 6.951, 2.883, and 1.461 at 20, 50,
and 100 lines, respectively. The corresponding ``reductions'' are negative:
$-595.1\%$, $-188.3\%$, and $-46.1\%$. Gemini first crosses below parity at
200 lines, where the ratio is 0.634 and the reduction is 36.6\%. It then reaches
0.194 at 800 lines and 0.128 at 1,600 lines before rising slightly to 0.139 at
2,000 lines.

Table~\ref{tab:size} gives the exact reductions. Break-even behavior is more
useful than the overall percentage for routing requests. In this configuration,
visual conversion is counterproductive for short Gemini contexts and becomes
competitive only at larger sizes. A harness needs a provider- and size-specific
rule rather than a global ``send code as images'' policy.

\begin{table*}[t]
  \centering
  \caption{Weighted input-token reduction by source length. Negative values
  mean that compact images receive more reported tokens than raw text.}
  \label{tab:size}
  \resizebox{\textwidth}{!}{\begin{tabular}{lrrrrrrrrr}
\toprule
Provider & 20 & 50 & 100 & 200 & 400 & 800 & 1,200 & 1,600 & 2,000 \\
\midrule
Anthropic & 60.3\% & 79.1\% & 83.4\% & 84.8\% & 86.9\% & 86.8\% & 86.8\% & 86.8\% & 86.8\% \\
OpenAI & 48.6\% & 66.5\% & 73.3\% & 77.7\% & 81.1\% & 80.9\% & 81.1\% & 81.1\% & 81.1\% \\
Gemini & -595.1\% & -188.3\% & -46.1\% & 36.6\% & 66.0\% & 80.7\% & 84.7\% & 87.2\% & 86.1\% \\
\bottomrule
\end{tabular}
}
\end{table*}

\subsection{RQ3: alias and workload heterogeneity}

Provider-level visual accounting is often shared across model aliases, but not
always. Three Anthropic models have the same 0.127 aggregate ratio, whereas
Claude Haiku 4.5 is higher at 0.165. All six OpenAI models return identical
aggregate text and image totals in this dataset, yielding 0.194 for each. Four
Gemini models share a 0.229 ratio, while Gemini 2.5 Flash is higher at 0.295.
These repeated values suggest common accounting rules within some families, but
the public responses do not reveal whether internal visual processing is
identical.

Weighted reductions are comparatively stable across the five language/file
prefixes. Anthropic ranges from 85.0\% for Go to 87.8\% for Rust. OpenAI ranges
from 78.7\% for Go to 83.3\% for JavaScript. Gemini ranges from 72.5\% for Java
to 78.5\% for Rust. Because each language corresponds to one file, these ranges
are descriptive workload differences, not language effects.

\begin{table}[t]
  \centering
  \caption{Weighted accounting by model. Repeated values identify shared
  \emph{reported} accounting in this dataset, not necessarily shared internal
  computation.}
  \label{tab:models}
  \scriptsize
  \resizebox{\columnwidth}{!}{\begin{tabular}{llrr}
\toprule
Provider & Model & Ratio & Reduction \\
\midrule
Anthropic & claude-fable-5 & 0.127$\times$ & 87.3\% \\
Anthropic & claude-haiku-4-5 & 0.165$\times$ & 83.5\% \\
Anthropic & claude-opus-4-8 & 0.127$\times$ & 87.3\% \\
Anthropic & claude-sonnet-5 & 0.127$\times$ & 87.3\% \\
Gemini & gemini-2.5-flash & 0.295$\times$ & 70.5\% \\
Gemini & gemini-3-flash-preview & 0.229$\times$ & 77.1\% \\
Gemini & gemini-3.1-flash-lite & 0.229$\times$ & 77.1\% \\
Gemini & gemini-3.1-pro-preview & 0.229$\times$ & 77.1\% \\
Gemini & gemini-3.5-flash & 0.229$\times$ & 77.1\% \\
OpenAI & gpt-5.3-codex & 0.194$\times$ & 80.6\% \\
OpenAI & gpt-5.4 & 0.194$\times$ & 80.6\% \\
OpenAI & gpt-5.5 & 0.194$\times$ & 80.6\% \\
OpenAI & gpt-5.6-luna & 0.194$\times$ & 80.6\% \\
OpenAI & gpt-5.6-sol & 0.194$\times$ & 80.6\% \\
OpenAI & gpt-5.6-terra & 0.194$\times$ & 80.6\% \\
\bottomrule
\end{tabular}
}
\end{table}

\begin{table}[t]
  \centering
  \caption{Weighted input-token reduction by language/file. Each row is one
  project file, so differences cannot be attributed to language alone.}
  \label{tab:languages}
  \begin{tabular}{lrrr}
\toprule
Language/file & Anthropic & OpenAI & Gemini \\
\midrule
Go & 85.0\% & 78.7\% & 74.8\% \\
Java & 86.1\% & 78.8\% & 72.5\% \\
Javascript & 87.7\% & 83.3\% & 78.4\% \\
Python & 85.4\% & 79.7\% & 73.1\% \\
Rust & 87.8\% & 81.5\% & 78.5\% \\
\bottomrule
\end{tabular}

\end{table}

\subsection{RQ4: page-sensitive accounting}

Most ratios decrease as source length grows because text tokens continue to
accumulate while image processing amortizes page-level overhead. The Gemini
2.5 Flash series nevertheless contains a sharp non-monotonic transition. We
repeated the Python image requests with modality-level usage metadata. At 800
lines, one rendered page received 2,322 image tokens; at 1,200 lines, two pages
received only 516 image tokens. For comparison, Gemini 3.5 Flash increased from
1,060 to 2,183 image tokens for the same page transition.

The repetition confirms the reported values for those requests but cannot
identify a mechanism. Resolution thresholds, model-specific media settings,
dynamic preprocessing, or service-side implementation changes are all possible.
We can conclude only that black-box image accounting need not be monotonic in
characters, pixels, or pages. A smooth compression rate can be misleading when
page boundaries are not inspected.

\section{Discussion}

\subsection{What the ratios do and do not imply}

The measurements support a precise statement: for this corpus and rendering
pipeline, sufficiently long compact-image requests receive substantially fewer
provider-reported input tokens than paired raw-text requests. DeepSeek-OCR's
97\% result shows that a specialized OCR system can recover document text with
high precision at moderate optical compression. Our measurements do not test
that property and therefore do not show lossless compression. Indentation
markers may be decoded incorrectly; small punctuation can disappear under
rasterization; multiple pages can disrupt cross-page structure; and a model may
summarize visible code without recovering the exact source needed for editing.

Likewise, fewer reported tokens do not imply proportional monetary savings.
Providers may price text and image tokens differently, change image accounting,
apply caching, or include fixed per-request charges. Nor do reported tokens
measure end-to-end latency: local rendering, PNG transfer, image encoding,
prefill, and decoding all contribute. The experiment intentionally avoids
converting ratios into dollar or compute claims.

\subsection{Implications for coding harnesses}

The data suggest a conditional architecture rather than a universal modality
switch. Small contexts should remain text, especially where an image minimum
dominates. Large, read-mostly context could be eligible for visual
representation if the task tolerates approximate recovery. Exact edit targets,
compiler errors, diffs, and user instructions should remain text-native until
fidelity is established. A mixed request could use images for broad repository
context and text for the active span.

Routing must also be calibrated per provider and rendering profile. A useful
policy would measure current break-even points, monitor image dimensions and
page counts, and fall back to text when an accounting or fidelity check fails.
Because aliases and service behavior can change, the calibration should be
versioned and rerun rather than hard-coded from this paper.

Visual conversion should additionally be compared against simpler mechanisms.
Retrieval can omit irrelevant files; textual compressors can preserve selected
tokens; prompt caching can avoid recomputing stable prefixes
\cite{jiang_2023_llmlingua,jiang_2024_longllmlingua,gim_2023_promptcache}.
Images become compelling only if they improve the joint frontier of usable
information, task quality, latency, and cost.

\subsection{Next experiment}

The next study should cross representation with task. Sample independent files
from multiple repositories and stratify by language, file length, nesting,
identifier density, and visual page count. For each file, compare raw text,
compact text without rasterization, readable code images, compact images, and a
textual compression baseline. Evaluate exact transcription, repository
retrieval, code question answering, summarization, defect localization, and
patch generation. Report syntax validity, tests passed, exact identifiers,
latency, bytes transferred, cached and uncached cost, and provider-reported
tokens.

Such a study would separate three effects currently combined here: compact
indentation, two-dimensional layout, and vision encoding. It would also connect
our accounting measurements to the quality-oriented evidence in CodeOCR,
LongCodeOCR, Glyph, and VTCBench
\cite{zhang_2026_codeocr,longcodeocr_2026,cheng_2026_glyph,vtcbench_2025}.

\section{Threats to Validity}

Empirical software-engineering guidance emphasizes construct validity,
measurement fairness, repetitions, and reproducibility
\cite{ralph_2021_empirical}. Our pilot is reproducible but intentionally narrow.

\paragraph{Construct validity.}
Provider-reported input tokens operationalize request accounting, not
information content, model compute, price, latency, or coding utility. The
phrase ``token reduction'' can invite all five interpretations, so we qualify it
throughout. The summarization instruction ensures valid multimodal requests but
does not make this a summarization benchmark because outputs are neither
normalized nor scored.

\paragraph{Internal validity.}
The treatment changes both modality and representation. Replacing indentation
spaces may reduce text length before rasterization and may affect readability;
without a compact-text arm, we cannot attribute the difference to pixels alone.
The study also lacks prompt-only and blank-image controls, so fixed request and
media overhead are inferred from curves rather than isolated experimentally.
Provider calls were made during a live service period rather than against frozen
local weights. Although paired inputs and prompts are fixed, undocumented
service routing or updates could alter counts. Requests were not repeatedly
sampled except for the targeted Gemini audit.

\paragraph{Conclusion validity.}
Nested prefixes are strongly dependent, and there is one file per language.
We do not use inferential statistics. Weighted ratios exactly summarize
the recorded requests but provide no estimate of a wider population. Pooling
aliases gives larger families more weight: OpenAI contributes six aliases,
Anthropic four, and Gemini five available aliases, although repeated token
signatures indicate only about five distinct accounting regimes. We report
providers separately and do not compute a cross-provider grand total.

\paragraph{External validity.}
The files come from prominent open-source projects but are not representative
of repositories, generated code, tests, notebooks, configuration, minified
files, or mixed-language contexts. Results may differ with font, syntax
highlighting, page width, image resolution, renderer, prompt, API region, and
model version. Lines of code are also a coarse size measure: the present run
does not stratify by bytes, characters, lexical tokens, nonblank lines, rendered
pixels, or page count.

\paragraph{Reliability and reproducibility.}
Revision-pinned URLs, exact line counts, scripts, lock files, raw JSON, paired
CSV, and deterministic regeneration reduce accidental variation. Validators
check duplicate keys and recompute ratios. However, reproducing identical
counts depends on access to proprietary APIs and on providers retaining their
current aliases and accounting behavior. Future replications should record
request timestamps, request identifiers, successful-call region, image hashes
and dimensions, and stable provider version identifiers where available.

\section{Artifact, Ethics, and AI Assistance}

The artifact includes corpus-fetching and rendering scripts, provider adapters,
resumable raw-result collection, data-integrity checks, deterministic figure and
table generation, a numeric-evidence registry, and a citation-verification lock.
API keys and locally fetched corpus/render files are excluded from version
control. Upstream source licenses and pinned revisions are documented. The
public repository is available at
\url{https://github.com/ron-42/code-image-token-accounting}. Its provider result
directories contain the complete model-alias $\times$ language/file $\times$
line-count matrix as raw JSON and paired CSV, including every successful text
and image observation rather than only the manuscript aggregates.

\subsection{Reproduction protocol}

Reproduction has four explicit stages. First, the corpus fetcher downloads each
pinned revision and emits the nested prefix files. A reviewer can compare hashes
and line counts before making any paid call. Second, the rendering wrapper
invokes the Node renderer from the repository root, deletes stale pages, and
fails if the renderer reports dropped characters. Page files are deterministic
inputs to the API stage, so they can be inspected independently of provider
access.

Third, each provider adapter traverses language, size, model, and modality in a
fixed order. Records are persisted atomically after every attempt. Resume mode
skips only successful keys, allowing transient failures to be retried without
repeating completed calls; force mode is required to replace a run. This
behavior matters because a partial matrix can otherwise silently mix old and
new service conditions. Replicators should archive the raw JSON immediately and
record collection time, region, SDK version, concrete model version when the
provider exposes one, request image dimensions, page count, and media-detail
settings.

Fourth, validation and analysis are separated from collection. The validator
constructs the unique key
\emph{provider/language/model/line-count/modality}, verifies model membership,
and checks each paired CSV row against its two raw JSON records. The analysis
then groups complete pairs and writes source CSV, LaTeX tables, a vector figure,
and a machine-readable registry mapping manuscript numbers back to selectors in
the generated data. The bibliography has a parallel lock containing official
DOI or arXiv metadata and a digest of each sentence-level support request.

\subsection{Reporting contract}

A replication should publish more than one aggregate percentage. At minimum it
should report: (1) the exact representation transform and renderer settings;
(2) source provenance and whether samples are independent; (3) complete and
failed model coverage; (4) the provider usage field used as the metric;
(5) weighted and unweighted estimands; (6) ratios by source size; and (7) page
dimensions and counts near every discontinuity. Negative reductions should be
shown as increases rather than described as savings. Cross-provider totals
should be avoided unless usage units and workload weights are explicitly made
comparable.

For claims about practical efficiency, token accounting is only one column of
the required report. A deployment-oriented replication must also include
task-quality metrics, end-to-end latency, bytes transferred, provider price
schedules, cache behavior, and local rendering cost. For claims about faithful
code transport, it should include exact transcription, identifier and literal
recovery, whitespace-sensitive syntax, and executable tests. This reporting
contract turns the current pilot's limitations into acceptance criteria for a
stronger study.

\subsection{Ethics and assistance}

The study uses public open-source code and does not involve human participants
or personal data. API use can incur cost and transmit source to external
providers; adopters should obtain authorization before applying the method to
private repositories.

AI coding and writing assistants supported implementation, literature
discovery, editing, and manuscript preparation. The author selected the study
scope, operated the experiments, inspected raw results, verified citations
against official records, and remains responsible for every claim. No
AI-generated output is treated as empirical evidence.

\section{Conclusion}

Compact rendered code can receive far fewer reported input tokens than raw
source text, but only after provider-specific overheads are amortized. In 675
paired requests, weighted reductions are 86.5\% for Anthropic, 80.6\% for
OpenAI, and 75.8\% for the five available Gemini aliases. Gemini reverses the
headline at small sizes and exhibits a reproduced page-sensitive
discontinuity, showing why aggregate compression percentages are insufficient.
The measurements motivate further evaluation but do not support deployment
claims. The remaining question is whether a visual representation preserves
enough exact and task-relevant code information to improve quality, latency,
and cost together in a coding system.

\balance
\bibliographystyle{ACM-Reference-Format}
\bibliography{references}

\end{document}